\documentclass[preprint,showpacs,preprintnumbers,amsmath,amssymb]{revtex4}

\usepackage{graphicx}
\usepackage{dcolumn}
\usepackage{bm}

\begin{document}

\preprint{APS/123-QED}

\title{Density--functional study of defects in two--dimensional circular nematic nanocavities}

\author{D. de las Heras}
\email{daniel.delasheras@uam.es}
\affiliation{Departamento de F\'{\i}sica Te\'orica de la Materia Condensada,
Universidad Aut\'onoma de Madrid,
E-28049, Madrid, Spain}

\author{L. Mederos}
\email{l.mederos@icmm.csic.es}
\affiliation{Instituto de Ciencia de Materiales de Madrid,
Consejo Superior de Investigaciones Cient\'{\i}ficas,
E-28049, Madrid, Spain}

\author{E. Velasco}
\email{enrique.velasco@uam.es}
\affiliation{Departamento de F\'{\i}sica Te\'orica de la Materia Condensada and
Instituto de Ciencia de Materiales Nicol\'as Cabrera,
Universidad Aut\'onoma de Madrid,
E-28049, Madrid, Spain}

\date{\today}

\begin{abstract}
We use density--functional theory to study the structure of two-dimensional defects inside a circular 
nematic nanocavity. The density, nematic order parameter, and director fields, as well as the 
defect core energy and core radius, are obtained in a thermodynamically consistent way
for defects with topological charge $k=+1$ (with radial and tangential symmetries) and
$k=+1/2$. An independent calculation of the fluid
elastic constants, within the same theory, allows us to connect with the local free--energy 
density predicted by elastic theory, which in turn provides a criterion to define a defect 
core boundary and a defect core free energy for the two types of defects. The radial and tangential 
defects turn out to have very different properties, a feature that a previous Maier--Saupe
theory could not account for due to the simplified nature of the interactions --which caused
all elastic constants to be equal. In the case with two $k=+1/2$ defects in the cavity, 
the elastic r\'egime cannot be reached due to the small radii of the cavities considered, but some
trends can already be obtained.
\end{abstract}

\pacs{Valid PACS appear here}
\maketitle

\section{Introduction}

The analysis of defects in liquid crystals is very important from many
points of view. In liquid--crystal applications, defects play a crucial role in
governing display--cell operation. Also, there are interesting theoretical
issues in different areas of physics concerning defects \cite{Trebin}, and
the stabilisation of defects has been observed and analysed in computer simulations
\cite{Chiccoli,Pelcovits,Lowen,Allen}.
A defect is a singularity in the director field
of the liquid crystal \cite{Kleman0,Lavrentovich}. Local properties of the liquid crystal, e.g. the
nematic order parameter, asymptotically relax to values of the bulk material
far from the singularity but, in its immediate neighbourhood, properties undergo 
abrupt (i.e. within molecular lengths) changes; this region somehow defines microscopically 
a boundary for the so--called defect core. 

Beyond the defect core, variations are smooth, so that the macroscopic elastic theory 
of Frank \cite{Frank} can be used, together with some assumptions about defect core energies 
and radii. Very often core energies are simply ignored. It would be desirable to have
estimations of these properties based on more microscopic approaches. In this context,
the Landau-de Gennes \cite{deGennes} theory has been extensively used to predict properties of
defects, but this theory is still mesoscopic in nature and makes no contact with
particle interactions. An alternative is to use computer simulations, but these are
generally time consuming for the study of defects. Therefore, the formulation of
theories based on molecular approaches are needed. A microscopic theory, of the Maier--Saupe
type, has been advanced \cite{RefWorks:36}, but it has some shortcomings; for example, 
it predicts all elastic constants to be equal, which causes different types of defects to have 
identical properties. This paper is devoted to exploring the consequences of another such
theories, namely a simple version of density--functional theory (DFT) for
hard anisotropic particles in two dimensions, which should give more realistic values for the
size and energies of defect cores since the theory predicts different
values for the elastic constants.

DFT is ideal to study liquid--crystal defects, since it self--consistently gives the 
thermodynamic and microscopic structural properties of the inhomogeneous nematic fluid. 
One advantage of DFT over traditional approaches is that elastic constants, in particular
the problematic surface
elastic constants, and other phenomenological parameters, do not appear explicitely in the theory,
but only implicitely through interactions and distribution functions in a free--energy
functional which is minimised (to all orders in the director spatial derivatives). The 
defect--core structure appears naturally, and this is ideal since, contrary to the usual
approximation within elastic theories of ignoring the defect cores in larger--sized nematic 
droplets, the contribution of defects cannot be ignored in nanocavities.
One question is why details of defects should be important to understand large-scale configurations of
the director field and defect motion. The microscopic approach enjoys some 
advantages whenever the relationship between bulk properties (such as elastic constants) and molecular structure 
and interaction parameters is required. Knowledge of the detailed structure of defects will not be crucial 
to understand large-scale configurations and defect motion in stable nematics subject to boundaries or in nematic 
matrices where colloidal particles are embedded, but there are circumstances where this may not be so. For example, 
in the kinetics of defect formation, re-organisation and anihilation, it may be important to know the 
core structure at short length scales. The microscopic approach can give useful estimates of 
free-energy changes, which are necessary to study the coarsening dynamics at a more microscopic level
using, for example, a relaxational dynamical equation. Also, the microscopic approach is essential 
at temperatures close to the clearing temperature, where defects act as nucleation seeds for the isotropic 
phase, and the structure and dynamics of the defects may be changing dramatically.

In DFT the structure of the fluid 
is summarised by the local density and orientational distribution functions, which in 
turn may be used to obtain the more familiar nematic order parameter and local director field; 
these two are basic to describe nematic fluids containing defects in the director field. The connection 
between the two descriptions is done via the local one--particle distribution
function, $\rho({\bm r},\phi)$, which gives the average number of particles at some position
${\bm r}$ with some orientation $\phi$ (on the two--dimensional plane). This quantity is obtained 
directly from DFT, and from it all interesting fields can be extracted, for example,
the microscopic director field $\hat{\bm n}({\bm r})$, which is obtained 
locally as the direction where the orientational part of the one--particle distribution function 
presents a maximum (the macroscopic nematic director could be obtained by some
coarse--grained average of the latter over some appropriate volume). 
Therefore the defect core region, along with the far neighbourhood of the singularity,
can be analysed within a single framework based on particle interactions. The 
computational demands of the method are high, however, and in the present paper
we restrict ourselves to the case of two--dimensional cavities of small radii
(in the nm scale). 

\begin{figure}
\includegraphics[width=3.4in]{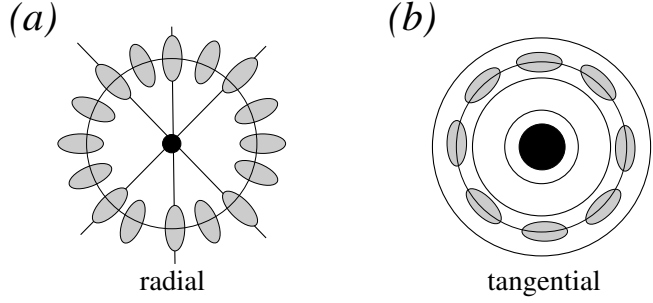}
\caption{\label{cores} Schematic of the two types of defects of topological 
charge $k=+1$ studied in this
work. (a) Radial defect, with particles pointing on average along the radial direction, which
excites splay distortion mode. (b) Tangential defect, with particles pointing on average along the 
tangential direction, which excites the bend distortion mode. Dark circular regions represent
the defect cores. Lines are tangent to the director field.}
\end{figure}

A defect is a singularity of the nematic
director field ${\bm n}$, characterised by a topological charge $k$, i.e. the number of turns
of the director when the singularity is completely encircled \cite{Kleman0}. Elastic theory 
assumes smooth spatial variations of the director and therefore is not able to account for the 
structure of the singularity.
In two dimensions the local elastic free--energy density can be written as 
\begin{eqnarray}
f_{\hbox{\tiny el}}({\bm r})=\frac{1}{2}k_1\left(\nabla\cdot\hat{\bm n}\right)^2+
\frac{1}{2}k_3\left|\hat{\bm n}\times\left(\nabla\times\hat{\bm n}\right)\right|^2,
\end{eqnarray}
where $k_1, k_3$ are elastic constants for splay and bend deformations (twist
deformations are not possible in 2D). Let us consider the two defects with topological 
charge $k=+1$ depicted in Fig. \ref{cores}, which are called `radial' (r) and `tangential' (t).
If $\phi$ is the polar angle of the position vector ${\bm r}$, then the director field
for the r defect is $\hat{\bm n}=(\cos{\phi},\sin{\phi})$, 
and $\nabla\cdot\hat{\bm n}=1/r$, $\nabla\times\hat{\bm n}={\bm 0}$, so that only
splay deformations are involved. In the t defect we have 
$\hat{\bm n}=(\sin{\phi},-\cos{\phi})$, $\nabla\cdot\hat{\bm n}=0$, 
$\left|\hat{\bm n}\times\left(\nabla\times\hat{\bm n}\right)\right|=1/r$, 
and the only deformations involved are
of bend type. A general deformation will involve both modes. Now, due to the singularity
at the origin (location of defect), the elastic free energy within an arbitrary area containing 
the origin will diverge logarithmically: elastic theory fails here, and it is necessary to
subtract this region by arbitrarily defining a core region, with free energy $F_n$ and radius
$r_n$. The free energy within a circle or radius $R$ will be:
\begin{eqnarray}
F_r&=&\pi k_1\log{\frac{R}{r_n^{(r)}}}+F_n^{(r)},\nonumber\\
F_t&=&\pi k_3\log{\frac{R}{r_n^{(t)}}}+F_n^{(t)}.
\end{eqnarray}
For $R\to\infty$ these energies diverge logarithmically, a situation that cannot arise
in practice due to the presence of defects with opposite charge in the material.

Little is known about the structure and properties of defect cores \cite{Lavrentovich}.
They are generally treated at a 
qualitative level, estimating the radius and defect--core energy in an approximate way
\cite{Lavrentovich}. Sometimes it is assumed these energies to be negligible compared with 
the elastic energy, and therefore defect cores are neglected altogether, a drastic simplification 
which can be severe if the system size is small. 
One of the first attempts to describe the defect core is due to Schopohl and Sluckin 
\cite{Schopohl}, who analysed a $1/2$--disclination using Landau--de Gennes theory
with the complete ordering tensor ${\bm Q}$. This study demonstrated that the core of
these defects does not consist of a region of isotropic material, but rather it is ordered
along the disclination line (a possibility that does not exist in 2D). Later Monte Carlo
simulations on a hard spherocylinder model by Hudson y Larson \cite{PhysRevLett.70.2916} 
corroborated this prediction, and also found a new structure with a stable triangular nucleus
for very elongated molecules. 

In the only truly microscopic theory presented so far,
Sigillo et al. \cite{RefWorks:36} used an approach
based on a Maier--Saupe theory with an orientational distribution function, analysing 
disclination lines of charge $k=+1$ within a cylinder. This is a three--dimensional setup,
while ours is a two--dimensional one. However, if one forgets about escape configurations,
the director field should in this case exhibit the same kind of configurations as in our
problem. The authors observed that the radius of the core decreases as the orientational order 
parameter increases. Also, they examined radial and tangential defects and analysed their
cores and their energies, obtaining that the two have the same size and energy. As the authors 
recognise this conclusion, which is certainly wrong, is due to 
the simplified interaction potential 
used, inherent in the Maier--Saupe theory, which predicts identical values for all the fluid
elastic constants.

Despite the reduced theoretical attention received, defect cores may play a very important 
role in many aspects of liquid--crystal science. Mottram \textit{et al.} 
\cite{disclination1997,RefWorks:34} studied disclination lines of charge $k=+1$ 
and $k=+1/2$ 
near the isotropic--nematic transition in 3D, and explained the impossibility of heating a
nematic material above a critical temperature $T_c>T_{\hbox{\tiny IN}}$ (clearing point) which is 
below the limit of metastability of the nematic phase, due to the growth of the isotropic core.
Defect cores properties may also be relevant in dynamical aspects such as defect motion
\cite{PhysRevLett.70.2916,PhysRevE.67.051705}.

In this paper we make a first attempt at calculating the properties of a defect core using
a microscopic approach based on density--functional theory, using hard--particle interactions.
One of our aims is to improve upon the results of Sigillo et al. \cite{RefWorks:36}
by making more sensible
predictions about the properties of the two types of defect cores investigated, namely with
radial and tangential symmetries, in a circular cavity. In Sec. \ref{theo} 
we briefly review the particle model and the DFT theory, together with the numerical
approach and the bulk behaviour. Sec. \ref{Elas} is devoted to the calculation of the elastic constants 
of the model. In contrast with elastic or Landau approaches, the DFT formalism does not require to specify 
which region is the core and which region is not the core, so that some criterion, similar to the
Gibbs dividing surface in the statistical mechanics of interfaces, is needed to define the
core. In order to analyse this problem it is necessary to compare the results from DFT with elastic
theory, and this demands knowledge of the splay and bend elastic constants $k_1$ and $k_3$. The values
of these constants can be obtained within the same DFT framework. In Sec. \ref{Results} we present
results for two types of point defects of charge $k=+1$ inside a circular cavity, placing 
emphasis on the size and energy of the defect core. Also, we discuss a configuration containing
two $k=+1/2$ defects, for which no definite conclusions can be drawn (due to the cavities explored 
being too small) but some trends can be obtained.

\begin{figure}
\begin{center}
\includegraphics[width=3.0in]{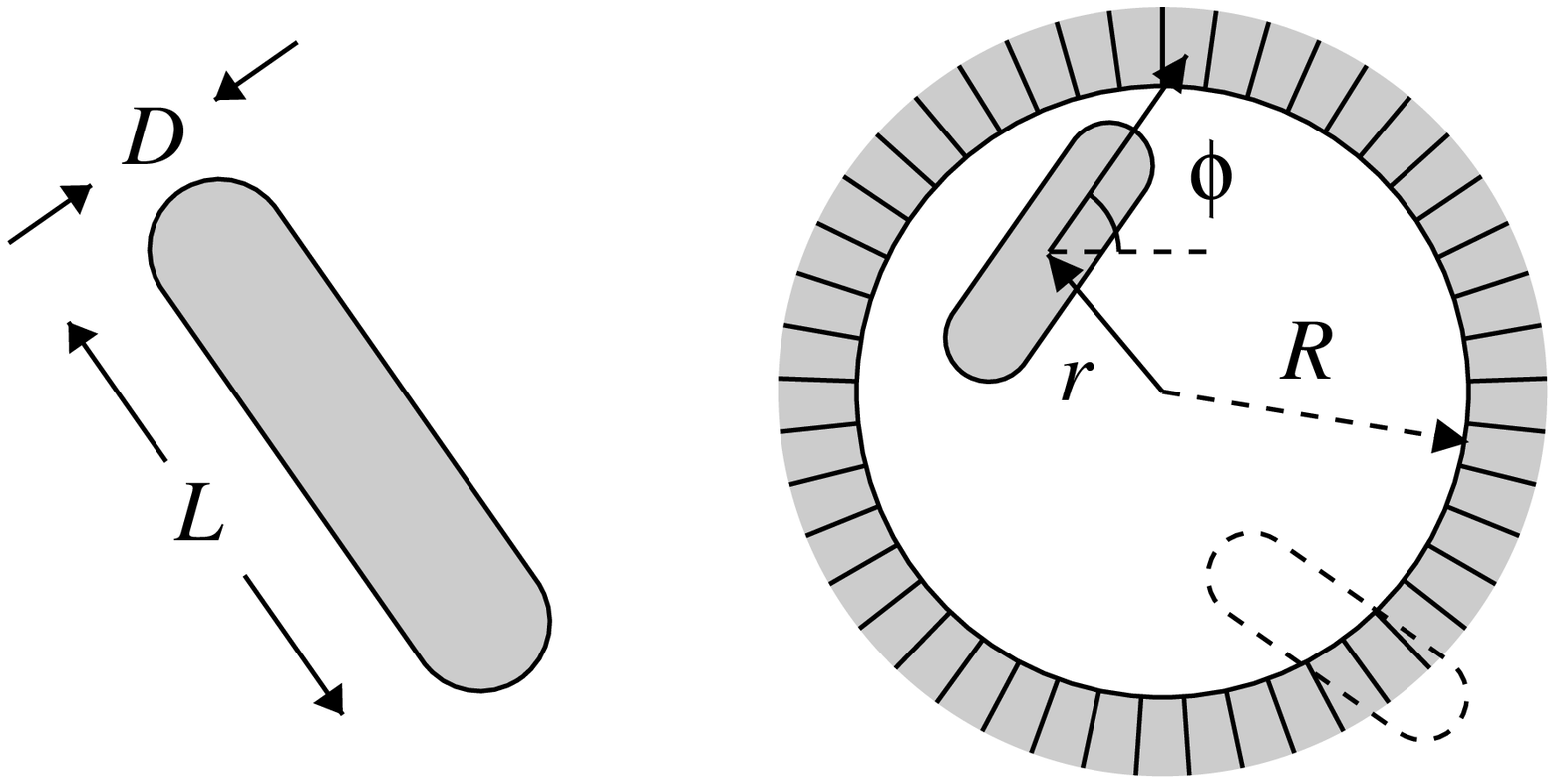}
\caption{Schematic of a hard discorectangle of total length $L+D$ and width $D$ (left), 
and cavity with a particle inside (right), showing
definition of radial distance $r$, polar angle $\phi$ and cavity radius $R$.
Dashed particle has its centre of mass right at the cavity wall and
cannot move further inside the wall (i.e. inside the cavity outer region).}
\label{2}
\end{center}
\end{figure}

\section{Theory}
\label{theo}

In a previous paper \cite{anterior} we have presented results for the structure, thermodynamics
and structural (Frederiks) transitions of nematics confined into two--dimensional circular 
nanocavities using DFT. Here we use the same version of the theory in the same setup, but with
an emphasis on defect core structure and energetics. Details of the theory were given 
in Ref. \cite{anterior}; here we give a summary of the main features.

The particle model used is the hard disco--rectangle (HDR), Fig. \ref{2}, which can be thought of as the
projection of a spherocylinder on a plane. A HDR particle has a rectangular section, 
of length $L$ and a diameter $D$, and two semicircular caps at the two ends of the rectangle,
also of diameter $D$. These particles interact via exclusion (i.e. configurations with
overlapping particles are not allowed, but particles are not interacting otherwise), 
and can form a two--dimensional
nematic at high volume fraction \cite{Frenkel}. As interactions are hard, the temperature dependence is
trivial, so the relevant intensive variable in the thermodynamics of this fluid will be the chemical 
potential (or, alternatively, the density).

In DFT one writes an approximate free--energy functional $F[\rho]$ in terms of the 
one--particle distribution function $\rho({\bm r},\phi)$, which can be split as
$\rho({\bm r},\phi)=\rho({\bm r})f({\bm r},\phi)$, where $f({\bm r},\phi)$
is the angular distribution function, and $\rho({\bm r})=\int d\phi\rho({\bm r},\phi)$
is the average local density. The free--energy functional is written as 
\begin{eqnarray}
F[\rho]=F_{\hbox{\tiny id}}[\rho] +F_{\hbox{\tiny exc}}[\rho] +F_{\hbox{\tiny ext}}[\rho]
\end{eqnarray}
with $F_{\hbox{\tiny id}}[\rho]$ the ideal contribution,
\begin{eqnarray}
\beta F_{\hbox{\tiny id}}[\rho]=\int_A d{\bm r}\rho({\bm r})\left\{\log{\left[\rho({\bm r})\Lambda^2-1\right]}-k^{-1} S_{\hbox{\tiny rot}}({\bm r})\right\},
\end{eqnarray}
where $A$ is the total area of the cavity, $\Lambda$ the thermal wavelength,
and $S_{\hbox{\tiny rot}}({\bm r})$ the local rotational entropy density:
\begin{eqnarray}
S_{\hbox{\tiny rot}}({\bm r})=-k\int_0^{2\pi} d\phi f({\bm r},\phi)\log{\left[2\pi f({\bm r},\phi)\right]}.
\end{eqnarray}
As usual, $\beta=1/kT$, $k$ being Boltzmann's constant.
The excess part, $F_{\hbox{\tiny exc}}[\rho]$, we write in terms of that of 
a reference fluid of locally parallel hard ellipses, which in turn is obtained exactly
from that of a hard--disc fluid. HDR, ellipses and discs will be chosen to have the same
particle area $v$ and, in the case of HDR and ellipses, the same aspect ratio. These
conditions are sufficient to fix $\sigma_{\parallel}$ and $\sigma_{\perp}$, the diameters
of the ellipses along the major and minor axes respectively, and from here $\sigma_e$, the
hard--disc diameter, with $\sigma_e^2=\sigma_{\parallel}\sigma_{\perp}$.
In the following we will use $\chi\equiv L/D=15$ (which gives $L=3.346\sigma_e$ and 
$D=0.223\sigma_e$). The excess free--energy per particle of the hard-disc fluid is
obtained from a theory due to Baus and Colot \cite{Baus}:
\begin{eqnarray}
\beta\Psi_{\hbox{\tiny exc}}(\eta)=(c_2+1)\frac{\eta}{1-\eta}+(c_2-1)\log{(1-\eta)},
\end{eqnarray}
where $c_2\simeq 0.1280$, $\eta=\rho_0 v$ is the packing (or volume) fraction, and
$\rho_0$ the mean number density. The excess free energy is then written as
\begin{eqnarray}
F_{\hbox{\tiny exc}}[\rho]=\int_A d{\bm r}\int_0^{2\pi}d\phi \rho({\bm r},\phi) \varphi({\bm r},\phi),
\label{F1}
\end{eqnarray}
where the local free--energy per particle is
\begin{eqnarray}
&&\varphi({\bm r},\phi)=\frac{\Psi_{\hbox{\tiny exc}}(\eta({\bm r}))}
{\pi\sigma_e^2\rho({\bm r})}\nonumber\\
&&\times
\int_A d{\bm r}^{\prime}\int_0^{2\pi}d\phi^{\prime}\rho({\bm r}^{\prime},\phi^{\prime})
v_{\hbox{\tiny exc}}({\bm r}-{\bm r}^{\prime},\phi,\phi^{\prime}).
\label{F2}
\end{eqnarray}
$v_{\hbox{\tiny exc}}$ is the overlap function of two HDR particles (equal to zero if
particles overlap and unity otherwise). This expression is a variation of 
the Parsons--Lee \cite{PL} theory for homogeneous fluids of hard rods, or (from a different
perspective) a variation of the Somoza--Tarazona \cite{Somoza} 
theory for inhomogeneous fluids of hard rods (both in three dimensions).
Finally, $F_{\hbox{\tiny ext}}[\rho]$ is the contribution from the external potential
(see Fig. \ref{2}):
\begin{eqnarray}
F_{\hbox{\tiny ext}}[\rho]=\int_A d{\bm r}\int_0^{2\pi}d\phi \rho({\bm r},\phi)v_{\hbox{\tiny ext}}
({\bm r},\phi).
\end{eqnarray}
The external potential acting on the particles will be chosen according to the type
of favoured particle orientation at the cavity surface. In the case of the radial defect it is
sufficient to use, as an external potential, a hard wall acting on the particle centres of mass: 
\begin{eqnarray}
v_{\hbox{\tiny ext}}({\bm r},\phi)=\left\{\begin{array}{ll}\infty,&r>R,\\\\0,&r<R,\end{array}\right.
\label{ext}
\end{eqnarray}
where $r$ is the radial distance measured from the centre of the cavity, 
and $R$ is the radius of the circular cavity. This choice is known to favour homeotropic
(i.e. perpendicular to the wall) orientation of the fluid director next to the wall \cite{anterior},
thus inducing an r--type defect. For the tangential defect a different choice is necessary 
(see Sec. \ref{tangential}).

The angular distribution function $f({\bm r},\phi)$ is parameterised
according to 
\begin{eqnarray}
f({\bm r},\phi)=\frac{\displaystyle e^{\alpha({\bm r})\cos{2\left[\phi-\psi({\bm r})\right]}}}
{\displaystyle \int_0^{2\pi} d\phi e^{\alpha({\bm r})\cos{2\phi}}},
\end{eqnarray}
where the field $\psi({\bm r})$ is the local tilt angle of the nematic
director, measured with respect to the $x$ axis, and
$\alpha({\bm r})$ is a variational function, related with the local nematic order
parameter $q({\bm r})$ by
\begin{eqnarray}
q({\bm r})=\int_0^{2\pi} d\phi f({\bm r},\phi)\cos{\left\{2\left[\phi-\psi({\bm r})\right]\right\}}.
\label{q}
\end{eqnarray}
These equations allow one to describe the configuration of the fluid by
means of the three local fields $\{\rho({\bm r}),q({\bm r}),\psi({\bm r})\}$; in
the following we will use the local packing fraction, $\eta({\bm r})=\pi\sigma_e^2\rho({\bm r})/4$, 
instead of the local density $\rho({\bm r})$, as basic density variable.
Finally, for a cavity of fixed radius $R$, we impose on the system a constant chemical
potential $\mu$, and minimise the cavity grand potential 
\begin{eqnarray}
\hat{\Omega}[\rho]=F[\rho]-\mu\int_A d{\bm r}\int_0^{2\pi}d\phi\rho({\bm r},\phi)
\end{eqnarray}
with respect to variations of the variables defined above.
To obtain the minimum the two dimensional space $xy$ is discretised into a square lattice 
with spacing $\Delta x=\Delta y=0.089\sigma_e$, with mesh points $(x_i,y_j)$, representing
40 points in a particle length $L+D$. The circular surface is approximated by
a zigzag line. The trapezoidal rule was used to calculate spatial integrations, while
angular integrals were approximated using Gaussian quadrature with $30$--$40$ roots.
The field variables $\{\eta({\bm r})$, $q({\bm r})$, $\Psi({\bm r})\}$
were discretised as $\eta_{ij}$, $q_{ij}$ and $\Psi_{ij}$, and the free--energy functional
was minimised using the conjugate--gradient method.

This model presents a bulk isotropic--nematic phase transition
for packing fraction $\eta_{\hbox{\tiny IN}}=0.257$ and reduced
pressure $pv_0/kT=0.98$ (estimates from simulation \cite{Frenkel}
give $\eta_{\hbox{\tiny IN}}=0.363$). The transition is of the
second order.

\section{Elastic constants}
\label{Elas}

In order to compare with elastic theory, we need some criterion to define the boundary of the 
defect core. In this respect it will be useful to compare the free--energy densities from DFT and 
elastic theory since, from this comparison, we can locate the boundary separating defect core 
from the outside region (where elastic theory should be valid) as the distance where both 
densities coincide. We will see that this definition is somewhat arbitrary, as it relies on
our definition of how close, numerically speaking, the two free--energy densities should be.
We come to this point later. For the moment, we note that the elastic free--energy density
contains elastic constants that we must be known in advance. These constants have to be calculated
within the same DFT scheme: the DFT free--energy density will smoothly tend to the value
predicted by elastic theory provided we use the values for elastic constants predicted by DFT. 
Then we calculate separately the elastic constants $k_1$ and $k_3$ in the framework of DFT. 
The expressions for the elastic constants are:
\begin{eqnarray}
k_1=-\frac{\Psi_{exc}(\eta_0)}{4\eta_0}\int_0^{2\pi} d\phi\int_0^{2\pi} d\phi'\rho'(\phi)\rho'(\phi'){\cal V}_{yy}(\phi,\phi'),\nonumber\\
k_3=-\frac{\Psi_{exc}(\eta_0)}{4\eta_0}\int_0^{2\pi} d\phi\int_0^{2\pi} d\phi'\rho'(\phi)\rho'(\phi'){\cal V}_{xx}(\phi,\phi'),\nonumber\\\label{ec6:k1k3}
\end{eqnarray}
where $\eta_0=\rho_0v$ is the bulk packing fraction. These constants are evaluated at the
uniform nematic (no spatial inhomogeneities or director distortions). 
The one-particle distribution function is then $\rho({\bm r},\phi)\equiv\rho(\phi)=\rho_0f(\phi)$. In the
expressions above, $\rho^{\prime}(\phi)$ is the derivative of the one-particle 
distribution function with respect to the tilt angle, $\rho^{\prime}(\phi)=
\partial\rho/\partial\psi=\rho_0\partial f/\partial\phi$, and where we defined
\begin{eqnarray}
{\cal V}_{ij}(\phi,\phi^{\prime})\equiv\int_{\hbox{\small excl. area}} d\mathbf{r}
v_{\hbox{\tiny exc}}(\mathbf{r},\phi,\phi^{\prime})x_ix_j.
\end{eqnarray}
The area integral over ${\bm r}$ is extended over the area of exclusion of two particles.
Details on how these expressions are obtained are to be found in the Appendix. An alternative and
equivalent way to obtain the elastic constants 
is to use the same confinement setup (circular cavity) defined above and 
impose a given director field with pure splay or bend deformations (Fig. \ref{cores}), setting the
density and nematic order parameters to the corresponding 
bulk values. Now if the free--energy density, as given
by evaluation of the functional, is represented along any one
of the cavity diameters (there is azimuthal symmetry), we can extract the elastic constants
by comparing with the radial dependence predicted by elastic theory in the intermediate region
(far from both the cavity centre and the cavity surface). We have seen already that the radial 
dependence is $\sim 1/r^2$ in both cases. Since no minimisation is implicit in this method, 
one can use very large cavities [$R\sim 100(L+D)$] so that the elastic constants can be obtained with accuracy. 

\begin{table} \begin{center} 
\begin{ruledtabular}
\begin{tabular}{ccccc}
$\Delta\mu/kT$ & $\eta_0$ & $q$ & $k_1/kT$ & $k_3/kT$ \tabularnewline \hline
$0.15$ & $0.270$ & $0.27$ & $0.08$ & $0.11$ \\
$0.75$ & $0.303$ & $0.63$ & $0.47$ & $1.13$ \\
$1.75$ & $0.360$ & $0.83$ & $0.93$ & $4.05$ \\
$2.75$ & $0.410$ & $0.90$ & $1.27$ & $8.72$ \\
$4.25$ & $0.470$ & $0.94$ & $1.70$ & $17.8$ \\
$5.25$ & $0.503$ & $0.96$ & $2.01$ & $25.2$ \\
$6.25$ & $0.533$ & $0.97$ & $2.38$ & $33.8$ \\
\end{tabular}
\end{ruledtabular}
\end{center} \caption{\label{I} Bulk properties of nematic fluid of HDR of aspect ratio $\chi=15$, as obtained from DFT. 
$\Delta\mu/kT$ is the excess chemical potential with respect to the isotropic--nematic 
coexistence value, in units of thermal
energy $kT$; $\eta_0$ and $q$ are the packing fraction and nematic order parameter; and $k_1/kT$, $k_3/kT$ are
values of elastic constants, also in units of thermal energy.}
\end{table}
 
\begin{figure}
\includegraphics[width=3.1in]{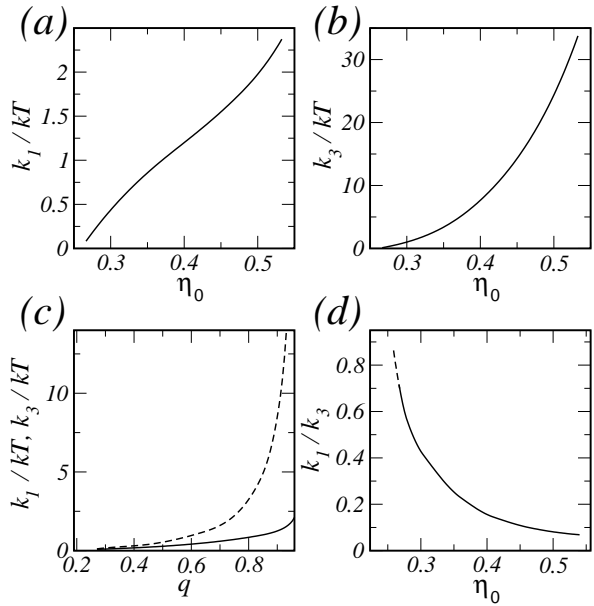}
\caption{\label{sc} Elastic constants of a nematic fluid of HDR particles with aspect ratio $\chi=15$, as obtained from DFT. (a) Splay elastic
constant $k_1$ in thermal energy units $kT$ as a function of packing fraction $\eta$. (b) Bend elastic constant $k_3$ in thermal energy units $kT$ as a function of packing fraction. (c) Splay (continuous line) and bend 
(dashed line) elastic constants
as a function of nematic order parameter $q$. (d) Ratio of elastic constants
as a function of packing fraction.}
\end{figure}

In Table \ref{I} and Fig. \ref{sc} we provide values for $k_1$ and $k_3$ (as expected, the two
strategies to obtain the elastic constants explained above give the same results, 
except for some tiny differences that come
from the numerical accuracy of angular and spatial integrals). The values of the elastic constants
are zero at the bulk transition. $k_3$ is always larger than $k_1$, and their difference increases
with density: when $\eta_0\approx0.4$ the difference is almost an order of magnitude, which
means that bend deformations are more costly energetically 
than splay deformations. This is an important point, as
most studies based on elastic theory assume the one--constant approximation $k_1=k_3$.
In our case (HDR particles with aspect ratio $\chi=15$) this approximation ceases to be valid
even very close to the isotropic--nematic transition [see Fig. \ref{sc}(d)]. In confined nematics 
under strong geometric restrictions such as the one studied here, nematic order is very frustrated
and stable nematic configurations are only obtained for conditions deep into the bulk nematic stability
region (i.e. and considerably far from the bulk transition); this means that the one--constant 
approximation will be very inaccurate. For particles with lower aspect ratios this problem will 
become less acute.

One consequence of this problem can be seen in the paper by Bates \cite{Bates}, where the nematic ordering
of hard spherocylinders lying on the surface of a sphere is examined via Monte Carlo simulation.
Geometry forces the creation of four defects of charge $+1/2$. However, analysis based on the one--constant
approximation predict that the defects are located at the vertices of a tetrahedron, while the simulations
show that they are in fact distributed along a great circle: in this way the director field arranges itself
in a way such that splay distortions are maximised, while bend distortions, much more costly energetically,
are minimised.

Another observation of our calculations concerns the bulk isotropic--nematic transition. This
transition has been studied by Bates and Frenkel \cite{Frenkel} by Monte Carlo simulation.
Assuming the transition to be of the Kosterlitz-Thouless type \cite{Stein}, and also that
the two elastic constants are equal, the transition should occur when the elastic constant reaches 
the critical value $k_c=8kT/\pi$. Using our values for the elastic constants and taking the
average $\bar{k}=(k_1+k_3)/2$, we obtain $\eta_{\hbox{\tiny IN}}=0.36$, in perfect agreement
with the simulations.

\section{Results}
\label{Results}

In this section we analyse various types of defects. We start with the radial configuration, r, where only
splay--type director distortions are present and there is a central defect of topological
charge $k=+1$. There follows the case of charge $k=+1$ but with a tangential, t, director field. 
Finally, we will consider point defects with charge $k=+1/2$.

\subsubsection{Radial defect with charge $k=+1$}

In Ref. \cite{anterior} we found that this defect
can only be stabilised at low chemical potential, close to the bulk isotropic--nematic transition.
As the chemical potential is increased, the r configuration becomes metastable and the central
$k=+1$ defect splits into two $k=+1/2$ defects. However, it is possible to impose the r configuration
by preparing the system so that the director is forced to always point radially,
keeping the director field unchanged during the conjugate--gradient minimisation. 

\begin{figure}
\begin{center}
\includegraphics[width=2.8in]{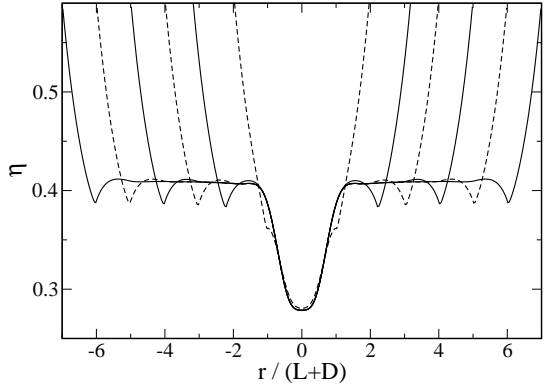}
\caption{For a radial defect of charge $k=+1$, local packing fraction $\eta$ as a 
function of radial distance $r$ from the cavity centre
(in units of particle length $L+D$), along an arbitrary cavity diameter, at relative chemical
potential $\Delta\mu/kT=2.75$. Lines correspond to different cavity radii: 
$R/(L+D)=1.98,3.17,3.98,4.98,5.98$ and $7.98$ (continuous and dashed lines alternate for a better
visualisation).}\label{fig6:convergencia}
\end{center}
\end{figure}

An example is given in
Fig. \ref{fig6:convergencia}, where the packing--fraction profile along one diameter is displayed.
The different cases shown correspond to increasing cavity radius, from $R/(L+D)=1.98$ to $7.98$.
Calculations are presented for fixed relative chemical potential $\Delta\mu/kT=2.75$, where
$\Delta\mu$ is referred to the value of $\mu$ at the bulk isotropic--nematic transition.
Three well--defined regions can be seen. In the central region a marked depletion in number of particles
is observed, which corresponds to the defect core. In the neighbourhood of this region the density is
quite constant, and as the inner surface of the cavity is approached a local minimum appears,
followed by a sharp density increase due to surface adsorption. Since we are mostly interested in the 
defect core, to minimise the effects of the surface on the core properties 
we need to consider as large a cavity as possible. Our present 
computational capabilities limit the radius of the cavity to $R\approx10(L+D)$. However, a simple 
inspection of the profiles seems to indicate that the surface effects are relatively weak, 
even in small cavities.
This figure clearly demonstrates that the size of the defect core is well defined even for cavities of
small radius [say $R\agt 3.2(L+D)$].

\begin{figure}
\begin{center}
\includegraphics[width=3.0in]{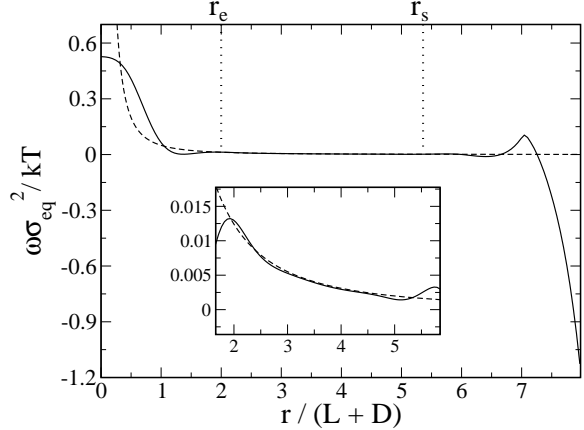}
\caption{Grand--potential density $\omega$, in units of $\sigma_e$ and $kT$, as a function of radial
distance (in units of particle length $L+D$),
along an arbitrary diameter of a cavity with $R=7.98(L+D)$ (continuous line).
The relative chemical potential is $\Delta\mu/kT=2.75$. The dashed line is the 
elastic free--energy density according to elastic theory. 
$r_e$ and $r_s$ are approximate radial distances for the boundaries of
the elastic region. The inset is a zoom of the central region.}
\label{fig6:densidadenergia}
\end{center}
\end{figure}

In the neighbourhood of the defect core there is a region dominated by elastic effects. If the defect
were very far from any surface this region would extend up to the surface, but the question is:
is it possible to obtain a truly elastic r\'egime for small cavities such as the ones investigated here?
To answer this question, we focus on the (grand--potential) 
free--energy density inside the cavity, $\omega(r)$, defined by
\begin{eqnarray}
\hat{\Omega}=\int_A d{\bm r} \omega(r).
\end{eqnarray}
In Fig. \ref{fig6:densidadenergia} the free--energy density is plotted as a function of radial distance
from the centre of the cavity, for a cavity radius  $R=7.98(L+D)$. The elastic free--energy density
$f_{\hbox{\tiny el}}(r)=k_1/2r^2$ is also included; to obtain this energy, the value for the $k_1$ elastic 
constant was taken from the DFT calculations (Table \ref{I}). Of course both free energies disagree 
in the central region of the cavity (where the free--energy 
density from elastic theory diverges at the singularity) and in the region close to the surface. 
However, there is an intermediate region, in the interval $r_e<r<r_s$,
[with $r_e\simeq 2.0-3.0(L+D)$ and $r_s\simeq 5.0-5.5(L+D)$] where 
the agreement is quite good; this is a signature of the elastic region. The conclusion that
an elastic r\'egime can indeed be defined was also reached by Sigillo et el. \cite{RefWorks:36} 
in their Maier--Saupe approach and indirectly in Landau--de Gennes approaches \cite{Schopohl}. 
The free--energy density may be used to loosely define a 
defect--core size in terms of the radial distance at which the free--energy density begins
to behave as $\sim r^{-2}$ (of course this is an ambiguous definition that, in practical terms, 
does not affect the numerical values of the defect--core properties significantly). In the case of Fig. 
\ref{fig6:densidadenergia} we obtain a size $2r_e\simeq 4-5(L+D)$; this should be a few times the
correlation length $\xi$, which is in agreement with calculations on three--dimensional 
defects by Landau-de Gennes theory \cite{Schopohl}.

\begin{figure}
\begin{center}
\includegraphics[width=3.1in]{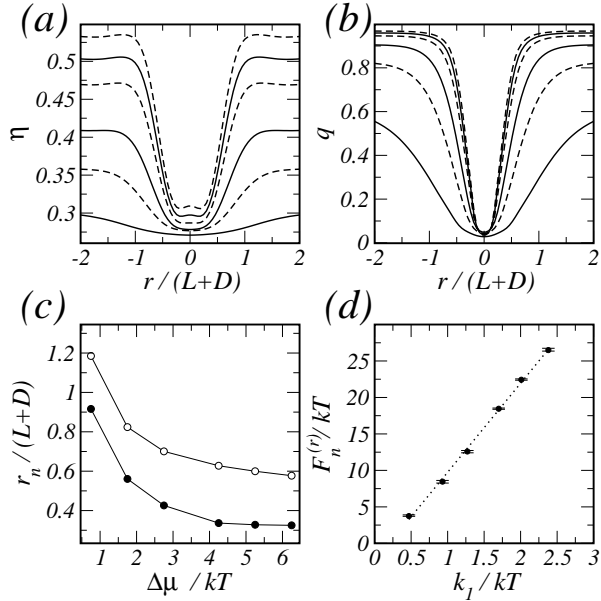}
\caption{Some properties of a nanocavity of radius $R=7.98(L+D)$.
(a) Local packing fraction $\eta$ as a function of radial distance $r$ from the cavity centre
(in units of particle length $L+D$) for various values of relative chemical potential: $\Delta\mu/kT=
0.75,1.75,2.75,4.25,5.25$ and $6.25$ (from bottom to top) and for a radial defect with $k=+1$. 
(b) Nematic order parameter $q$
as a function of radial distance $r$ from the cavity centre
(in units of particle length). Profiles as in panel (a). (c) Core radius $r_n$ in units of particle
length as a function of relative chemical potential. Filled circles: $r_n^{(1)}$. Open circles: 
$r_n^{(2)}$. 
(d) Core energy $F_n^{(r)}$ as function of splay elastic constant $k_1$, both in thermal energy units. 
Error bars were calculated with the two choices $r_e=2.5(L+D)$ and $3.0(L+D)$ for the
upper limit in the integral of Eqn. (\ref{Fn}).
The straight line is a linear fit.}\label{nucleo}
\end{center}
\end{figure}

The properties of a cavity of radius $R=7.98(L+D)$ are summarised in Fig. \ref{nucleo}.
The profiles of the nematic order parameter, Figs. \ref{nucleo}(a) and (b), indicate that
the core radius decreases as the chemical potential $\mu$ increases. To quantify this 
effect more precisely and analyse the depletion of the order parameter that occurs inside the 
central region, we have defined two additional measures of the defect--core radius,
$r_n^{(1)}$ and $r_n^{(2)}$, as the inflection points in the nematic order parameter and the density 
profiles, respectively; these two quantities do not coincide with, but should behave 
like, the energy--based measure $r_e$ as thermodynamic conditions are varied).
In Fig. \ref{nucleo}(c)
we plot these quantities as symbols. Both have a similar behaviour: they decrease quickly with $\mu$
and saturate at high chemical potential, with the inflection point of the nematic order parameter
saturating a bit earlier.

The fact that the core radius decreases with $\mu$ does not mean that its effects propagate to a 
smaller region; in fact, the result is quite 
the opposite. The difference between bulk and core densities increases
with $\mu$, while the intermediate region extends to larger distances. However, the effective core 
radius $r_e$ (where the free--energy density differs significantly from the elastic one) is usually 
in the interval $2.5-3(L+D)$, largely independent of $\mu$ for large $\Delta\mu$.

The successful identification of an asymptotic elastic region and the ensuing possibility of defining a 
defect--core boundary allows us to associate a free energy $F_n$ with the defect core. 
This we do by integrating the excess of grand--potential density over a uniform fluid at the same
chemical potential inside a circle of radius $r_e$ (where elastic behaviour sets in): 
\begin{eqnarray}
F_n=2\pi\int_0^{r_e} dr r \omega(r).
\label{Fn}
\end{eqnarray}
This energy is represented in Fig. \ref{nucleo}(d) as a 
function of the elastic constant $k_1$. The calculation has been done using $r=2.75(L+D)$ as a
cut--off distance, but calculations were also done using $2.5(L+D)$ and $3.0(L+D)$ to see
the effect of changing the cut--off; error bars in the data correspond to these two limits.
As can be seen in the figure, the differences are very small. 

In phenomenological treatments it is usual to assume that the free energy of a disclination core of charge
$k$ is $F_n=k^2\pi\bar{k}$, where $\bar{k}$ is an elastic constant \cite{Lavrentovich}
in the one--constant approximation
$\bar{k}=k_1=k_3$) which, for the radial defect, is $k_1$. Our DFT results give support to the linear 
relation between $F_n^{(r)}$ and $k_1$, but the slope (obtained by a linear fit) is equal to $12.2$,
which is four times larger than that predicted by the phenomenological theory for a point defect of
charge $k=+1$. The dependence of $F_n^{(r)}$ on chemical potential is also linear, with a
slope of $4.09$ (not shown).

\subsubsection{Tangential defect with charge $k=+1$}
\label{tangential}

In this case the director field only supports bend distortions, as shown in Fig. \ref{cores}.
To stabilise such a structure we need a surface potential that favours planar anchoring, i.e.
particle orientations tangential to the surface. We use the following model for external potential:
\begin{eqnarray}
v_{\hbox{\tiny ext}}(r,\phi)=\left\{\begin{array}{ll}
\infty,&r>R,\\\\
V_0\cos{2(\phi-\psi)}e^{-\alpha (R-r)},&r<R,\end{array}\right.
\label{c6:Vext}
\end{eqnarray}
where $V_0$ is the surface strength. For $V_0$ large enough, the surface favours tangential
anchoring; we have checked that this is the case, e.g. for $V_0=0.7kT$ and $\alpha=1.08(L+D)^{-1}$. 
These are the values we will be using in the following to study a defect with tangential anchoring.

\begin{figure}
\begin{center}
\includegraphics[width=3.2in]{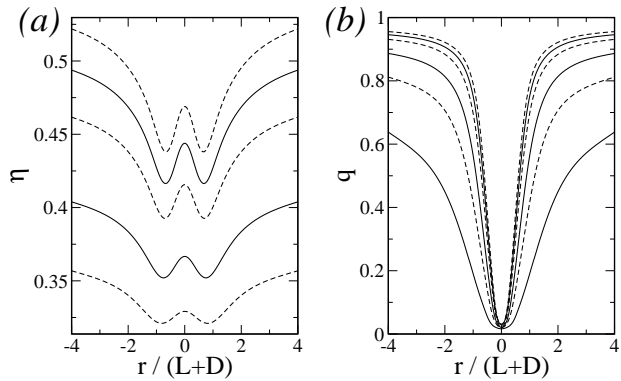}
\caption{(a) Local packing fraction $\eta$ as a function of radial distance $r$ from the cavity centre
(in units of particle length $L+D$) for various values of relative chemical potential: $\Delta\mu/kT=
0.75,1.75,2.75,4.25,5.25$ and $6.25$ (from bottom to top) and for a tangential defect with $k=+1$. 
(b) Nematic order parameter $q$
as a function of radial distance $r$ from the cavity centre
(in units of particle length). Profiles as in panel (a). All data pertain to the case $R=7.98(L+D)$.} 
\label{bend}
\end{center}
\end{figure}

The inclusion of an external field with an exponential decay means that the surface interacts
with the fluid at longer distances than in the previous case. An additional feature is that, 
since the splay elastic constant is smaller than the bend elastic constant, the
size of the defect core is larger. Both these effects play against the possibility of
reaching the elastic r\'egime in the region between the defect and the surface. Therefore, much
larger cavities are needed. Our computational limit is $R\sim 15(L+D)$, which is not large enough
to obtain reasonably accurate estimates of the core energy, for example. The only safe conclusion
is that this energy is significantly larger than that of the radial defect.

Despite this problem, it is instructive 
to study the structure of the defect core in a qualitative way. Fig. \ref{bend}
shows the order parameter profiles in the defect--core region for a cavity of
radius $R=7.98(L+D)$ (for larger cavities the profiles will be slightly different). 
The nematic order parameter behaves similarly
as in the previous case, save the different size. The density has a pronounced maximum
at the core centre. The size of the central depleted region is larger than one
particle length, allowing for a higher particle concentration inside the defect core.

\subsubsection{Defect with charge $k=+1/2$}

\begin{figure}
\begin{center}
\includegraphics[width=3.2in]{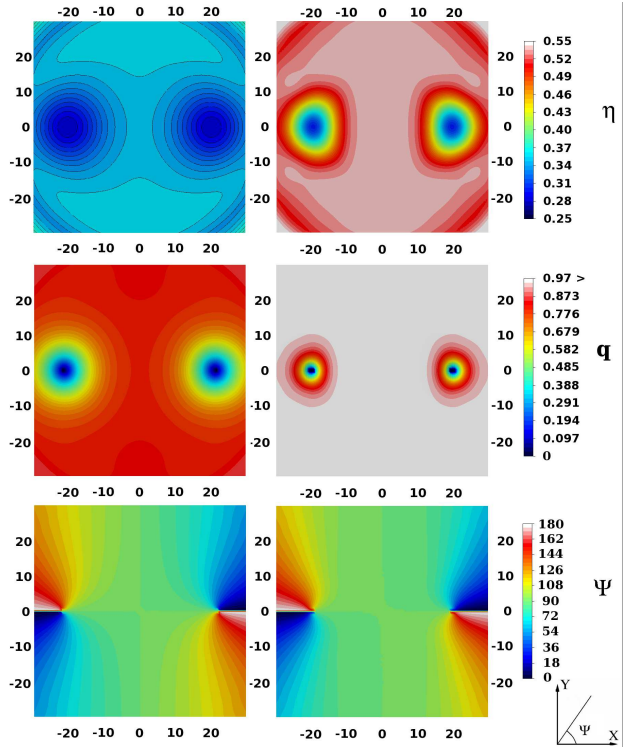}
\caption{Contour plots with respect to $xy$ coordinates
for local packing fraction $\eta$ (upper row),
nematic order parameter $q$ (middle) and director tilt angle $\Psi$ (lower) for two
configurations with relative chemical potentials $\Delta\mu=1.75 kT$ (left column)
and $\Delta\mu/kT=6.75$ (right column) in a cavity of radius $R=3.18(L+D)$.
$xy$ coordinates in units of $D$, and tilt angle is given in degrees.}
\label{3dim2}
\end{center}
\end{figure}

The present geometry can also be used to explore a more interesting case: a defect
with charge $k=+1/2$. The minimum--energy state contains two defects of charge $k=+1/2$,
separated by a distance $d_0$, when the cavity radius is sufficiently large, but the analysis
is more complicated here, as two additional minimisations are required: a partial one with
respect to the `fast' variables at fixed defect separation, and a minimisation with respect 
to the defect separation $d_0$, which is a slow variable. As a result, the
computation time increases by an order of magnitude. The practical consequence is that the 
maximum radius of the cavity that can be analysed is reduced, and the task of splitting contributions
of defect cores from the rest becomes harder.

In Fig. \ref{3dim2} we have plotted the order parameters for a configuration with two
$k=+1/2$ defects, at two different chemical potentials. The configurations were obtained by 
minimising the functional in a cavity of radius $R=3.18(L+D)$ 
(only a region of size $30\times 30D^2$ containing the two defect 
cores is shown). In the left column local packing fraction $\eta$ (top), nematic
order parameter (middle) and tilt angle (bottom) are shown for the case $\Delta\mu=1.75kT$. 
The right column shows the same profiles when the chemical potential is increased to 
$\Delta\mu=6.75kT$. In the corners of the density plots the structure has radial symmetry: 
this is a surface effect. This is an indication that larger cavities may be necessary for a
more detailed study. We can clearly see that the core size decreases considerably as the 
chemical potential is increased (this effect is more visible in the nematic order parameter). 
Another remarkable effect is the loss of radial symmetry of the defect as the chemical potential
is increased. The profiles in the left column (low chemical potential) have an almost radial
symmetry with respect to the defect core (save the tilt angle, obviously).
As $\mu$ is increased (right column), the core shrinks in all directions, especially along
the direction joining the two defects, where the director field is constant. 

\begin{figure}
\begin{center}
\includegraphics[width=3.2in]{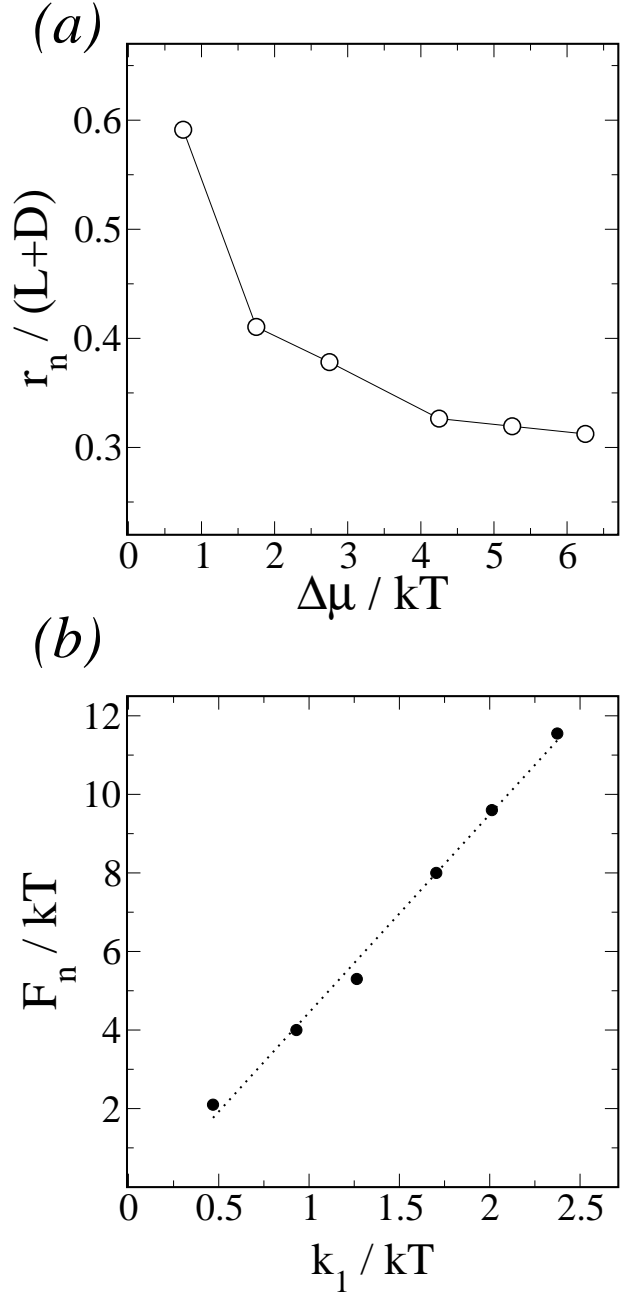}
\caption{(a) Radius $r_n$ as a function of relative chemical potential $\Delta\mu$, and (b)
core energy $F_n$ as a function of elastic constant $k_1$ for a defect of charge $k=+1/2$.}
\label{defectos12}
\end{center}
\end{figure}

Unfortunately, we have not been able yet to study cavities large enough for the core structure
and the surface structure to relax completely to the elastic limit, but qualitative estimates
can be obtained for the relevant properties of the core. This can be seen in Fig. \ref{defectos12}. 
In panel (a) we plot the average radius of a defect core as a function of the chemical potential. 
Similar to the radial defect, the average radius $r_n$ has been defined as the inflection point of the 
density profile, averaged over all directions (since here there is no angular symmetry).
We can see that there is a rapid decay as $\mu$ increases, and $r_n$ levels off at a value
approximately equal to half the value for the radial defect [see Fig. \ref{nucleo}(c)].
Therefore, the core area, which is proportional to $r_n^2$, is about four times less in the
case $k=+1/2$ than in the case $k=1$. In panel (b) of the same figure, we plot the
defect core energy as a function of the elastic constant $k_1$. The straight line is a
linear regression with slope $m=5.04$, approximately a factor $2.5$ smaller than in the 
$k=+1$ defect. We could expect a factor 2 beforehand, since we know in advance that,
even in very small cavities, the structure with two $k=+1/2$ defects is more stable than 
that with a single, radial defect. In this case the largest contribution to the free energy
comes from the defect cores, and therefore the energy of both cores plus their repulsive energy
must be at most equal to the energy of the radial defect. We mentioned before that the defect 
energy is generally taken to be $F_n=k^2\pi\bar{k}$ and we would expect a factor $1/4$ in the case $k=+1/2$
with respect to the radial defect (the size is approximately four times smaller). The behaviour
of the director field in the core region exhibits bend--like distortions 
($\mathbf{\nabla\times n}\neq 0$) for $k=+1/2$, which do not appear in the radial defect and
can be the origin of such a difference.

\section{Summary and conclusions}

In summary, we have studied the core properties of a defect in 2D, using a 
DFT (microscopic) model, free from fitting parameters, that includes consistently variations
in density and nematic order parameter. All cases studied predict the formation of
an isotropic region in the defect, something to be expected in 2D. The core free energy
is proportional to the elastic constant, with a varying proportionality constant that
depends on the type of defect studied; this is due to the different energetic
cost associated with deformations of splay and bend type. The size of the defect cores is 
on the order of a few particle lengths, and decreases as the nematic ordering of the
surrounding fluid increases. The core size saturates for strong nematic ordering.

The study of a 2D defected nematic fluid, used here mainly for computational reasons, may be useful to
understand 3D phenomena in the physics of defects. As mentioned in the introduction, knowledge on the
structure and energetics of defects may be important in dynamical problems, such as defect formation or
nucleation and coarsing of the nematic and isotropic phases. This structure may be changing in time and
a microscopic approach may be helpful to follow the dynamics via relaxation equations that involve
the gradient of a free energy. A natural extension of our work therefore involves the study of 3D nematic 
fluids and their defects. Schemes where the microscopic and mesoscopic approaches are combined would
also be useful in the above-mentioned problems, and work along this avenue is under way in our group. 

\begin{acknowledgments}
We acknowledge financial support from Ministerio de Educaci\'on y Ciencia (Spain) under
Grant Nos. FIS2008-05865-C02-02, FIS2007-65869-C03-C01, FIS2008-05865-C02-01, 
and Comunidad Aut\'onoma de Madrid (Spain) under Grant No. S-0505/ESP-0299.
\end{acknowledgments}

\appendix
\section{Calculation of elastic constants}

Expressions for the elastic constants of a 3D nematic liquid crystal were derived
by Poniewierski and Stecki \cite{Ponie} using a direct correlation function route.
>From these expressions it is easy to write the corresponding 2D expressions. Obtaining
the direct correlation function of the theory (which, in our Onsager-type theory, is basically 
the Mayer function) one can obtain explicit expressions in terms of integrals over
the excluded area and the orientational distribution functions. Here we present an
alternative derivation, valid only in 2D, in terms of expansions in the local tilt angle
$\psi({\bm r})$. We start from Eqns. (\ref{F1}) and (\ref{F2})
for the excess free energy of a nematic with constant density, 
expressed explicitely in terms of the tilt angle:
\begin{eqnarray}
&&F_{\rm exc}[\rho]=\frac{\Psi_{\rm exc}(\eta_0)}{4\eta_0}
\int d{\bm r}\int d\phi\rho[\phi-\psi({\bm r})]\nonumber\\\nonumber\\&&\times
\int\!\!\int d{\bm r}^{\prime}
d\phi^{\prime}v_{\hbox{\tiny exc}}({\bm r}-{\bm r}^{\prime},\phi,\phi^{\prime})
\rho[\phi^{\prime}-\psi({\bm r}^{\prime})]
\end{eqnarray}
Now we expand the second local density
$\rho(\phi^{\prime}-\psi({\bm r}^{\prime}))$ in
$\psi({\bm r}^{\prime})$ around ${\bm r}^{\prime}={\bm r}$.
Letting $\Delta{\bm r}={\bm r}^{\prime}-{\bm r}$, we have:
\begin{eqnarray}
\psi({\bm r}^{\prime})=\psi({\bm r})+\Delta{\bm r}
\cdot\nabla_{\bm r}\psi+\frac{1}{2}\left[
\Delta{\bm r}\cdot\nabla_{\bm r}\right]^2\psi+\cdots
\nonumber\\
\label{3}
\end{eqnarray}
Then we expand the density:
\begin{eqnarray}
\rho[\phi^{\prime}-\psi({\bf r}^{\prime})]&=&
\rho[\phi^{\prime}-\psi({\bf r})]+
\left.\frac{\partial\rho}{\partial\psi}\right|_{\bf r}
\left[\psi({\bf r}^{\prime})-\psi({\bf r})\right]\nonumber\\\nonumber\\&+&
\frac{1}{2}\left.\frac{\partial^2\rho}{\partial\psi^2}\right|_{\bf r}
\left[\psi({\bf r}^{\prime})-\psi({\bf r})\right]^2+...
\end{eqnarray}
Substituting (\ref{3}) and keeping terms up to third order in the gradient:
\begin{eqnarray}
&&\rho[\phi^{\prime}-\psi({\bm r}^{\prime})]=
\rho(\phi^{\prime}-\psi({\bm r}))+
\left.\frac{\partial\rho}{\partial\psi}\right|_{\bm r}\nonumber\\\nonumber\\&&\times
\left(\Delta{\bm r}\cdot\nabla_{\bm r}\psi+
\frac{1}{2}\left(\Delta{\bm r}\cdot\nabla_{\bm r}\right)^2\psi+
O\left(\nabla_{\bm r}\psi\right)^3\right]
\nonumber\\\nonumber\\&&+
\frac{1}{2}\left.\frac{\partial^2\rho}{\partial\psi^2}\right|_{\bm r}
\nonumber\\\nonumber\\&&\times
\left[\Delta{\bm r}\cdot\nabla_{\bm r}\psi+
\frac{1}{2}\left(\Delta{\bm r}\cdot\nabla_{\bm r}\right)^2\psi+
O\left(\nabla_{\bm r}\psi\right)^3\right]^2+...
\nonumber\\\nonumber\\&&=
\rho[\phi^{\prime}-\psi({\bm r})]+
\rho^{\prime}_{\psi}
[\phi^{\prime}-\psi({\bm r})]\Delta{\bm r}
\cdot\nabla_{\bm r}\psi+
\nonumber\\\nonumber\\&&+\frac{1}{2}
\rho^{\prime}_{\psi}[\phi^{\prime}-\psi({\bm r})]
\left(\Delta{\bm r}\cdot\nabla_{\bm r}\right)^2\psi\nonumber\\\nonumber\\&&+
\frac{1}{2}
\rho^{\prime\prime}_{\psi}[\phi^{\prime}-\psi({\bm r})]
\left(\Delta{\bm r}\cdot\nabla_{\bm r}\psi\right)^2+...
\end{eqnarray}
The first term gives the free energy of the undistorted nematic 
(since $\psi({\bm r})$ is supposed to be a slowly varying field). The elastic
free energy is then:
\begin{eqnarray}
&&F_{\rm el}[\rho]=
\frac{\Psi_{\rm exc}(\eta_0)}{8\eta_0}\nonumber\\\nonumber\\&&\times
\int\!\!\int d{\bm r}
d\phi\rho(\phi-\psi({\bm r}))
\int\!\!\int d{\bm r}^{\prime}
d\phi^{\prime}v_{\hbox{\tiny exc}}({\bm r}^{\prime},\phi,\phi^{\prime})
\nonumber\\\nonumber\\&&\times
\left\{\rho^{\prime}_{\psi}
[\phi^{\prime}-\psi({\bm r})]{\bm r}^{\prime}
\cdot\nabla_{\bm r}\psi+
\frac{1}{2}
\rho^{\prime}_{\psi}[\phi^{\prime}-\psi({\bm r})]
\left[{\bm r}^{\prime}\cdot\nabla_{\bm r}\right]^2\psi
\right.
\nonumber\\\nonumber\\&&+\left.
\frac{1}{2}
\rho^{\prime\prime}_{\psi}(\phi^{\prime}-\psi({\bm r}))
\left[{\bm r}^{\prime}\cdot\nabla_{\bm r}\psi\right]^2.
\right\}
\end{eqnarray}
We can take $\psi({\bm r})=0$ in the argument of the density profiles 
and its derivatives; the elastic free--energy density is then:
\begin{eqnarray}
&&f_{\rm d}({\bm r})=\frac{\Psi_{\rm exc}(\eta_0)}{8\eta_0}
\int d\phi\rho(\phi)
\int\!\!\int d{\bm r}^{\prime}
d\phi^{\prime}v_{\hbox{\tiny exc}}({\bm r}^{\prime},\phi,\phi^{\prime})
\nonumber\\\nonumber\\&&\times
\left\{\rho^{\prime}_{\psi}
(\phi^{\prime})\left({\bm r}^{\prime}
\cdot\nabla_{\bm r}\psi\right)+\frac{1}{2} \rho^{\prime}_{\psi}(\phi^{\prime})
\left({\bm r}^{\prime}\cdot\nabla_{\bm r}\right)^2\psi\right.\nonumber\\\nonumber\\
&&\left.+\frac{1}{2}
\rho^{\prime\prime}_{\psi}(\phi^{\prime})
\left({\bm r}^{\prime}\cdot\nabla_{\bm r}\psi\right)^2
\right\}
\end{eqnarray}
Now we note that
\begin{eqnarray}
\int d{\bm r}^{\prime}v_{\hbox{\tiny exc}}({\bm r}^{\prime},\phi,\phi^{\prime})
{\bm r}^{\prime}={\bm 0},
\end{eqnarray}
due to the symmetry $V({\bm r},\phi,\phi^{\prime})=
V(-{\bm r},\phi,\phi^{\prime})$, and the term linear in the
gradient of $\psi({\bm r})$ vanishes, as it should be.
Defining the dyadic
\begin{eqnarray}
\tilde{{\cal V}}(\phi,\phi^{\prime})\equiv
\int d{\bm r}v_{\hbox{\tiny exc}}({\bm r},\phi,\phi^{\prime}){\bm r}{\bm r},\hspace{0.3cm}
\end{eqnarray}
we get
\begin{eqnarray}
&&F_{\rm el}[\rho]=\frac{\Psi_{\rm exc}(\eta_0)}{8\eta_0}
\nonumber\\\nonumber\\&&\times
\sum_{\beta\gamma}\int\!\!\int d\phi d\phi^{\prime}
{\cal V}_{\beta\gamma}(\phi,\phi^{\prime})
\int d{\bm r}\rho(\phi-\psi({\bm r}))
\nonumber\\\nonumber\\&&\times
\left\{\rho^{\prime}_{\psi}[\phi^{\prime}-\psi({\bm r})]
\partial_{\beta\gamma}\psi({\bm r})
+\rho^{\prime\prime}_{\psi}[\phi^{\prime}-\psi({\bm r})]
\partial_{\beta}\psi({\bm r}) \partial_{\gamma}\psi({\bm r}) 
\right\}.\nonumber\\
\label{A9}
\end{eqnarray}
Now we integrate the term with the second derivatives by parts:
\begin{eqnarray}
&&\int d{\bm r}\rho[\phi-\psi({\bm r})]
\rho^{\prime}_{\psi}[\phi^{\prime}-\psi({\bm r})]
\partial_{\beta\gamma}\psi({\bm r})\nonumber\\\nonumber\\&&=
\left.\rho[\phi-\psi({\bm r})]
\rho^{\prime}_{\psi}[\phi^{\prime}-\psi({\bm r})]\partial_{\gamma}\psi({\bm r})
\right|_{x_{\beta}=\hbox{const.}}
\nonumber\\\nonumber\\&&-
\int d{\bm r}\left\{\rho^{\prime}_{\psi}[\phi-\psi({\bm r})]
\rho^{\prime}_{\psi}[\phi^{\prime}-\psi({\bm r})]\right.\nonumber\\\nonumber\\&&\left.+
\rho[\phi-\psi({\bm r})]
\rho^{\prime\prime}_{\psi}[\phi^{\prime}-\psi({\bm r})]\right\}
\partial_{\beta}\psi({\bm r})\partial_{\gamma}\psi({\bm r}).
\end{eqnarray}
The $\rho\rho^{\prime\prime}$ terms in (\ref{A9}) cancel out, and the surface term
is neglected. Then:
\begin{eqnarray}
&&f_{\rm el}({\bm r})=-\frac{\Psi_{\rm exc}(\eta_0)}{8\eta_0}
\sum_{\beta\gamma}\int\!\!\int d\phi d\phi^{\prime}
{\cal V}_{\beta\gamma}(\phi,\phi^{\prime})\nonumber\\\nonumber\\
&& \rho^{\prime}_{\psi}(\phi-\psi({\bm r}))
\rho^{\prime}_{\psi}(\phi^{\prime}-\psi({\bm r}))
\partial_{\beta}\psi({\bm r}) \partial_{\gamma}\psi({\bm r}),
\end{eqnarray}
which is the searched--for expression. In terms of the unit vector
$\hat{\bm\omega}$ along the particle axis, we have
$\hat{\bm\omega}\cdot\hat{\bf n}=\cos{(\phi-\psi)}$, and
\begin{eqnarray*}
&&\rho^{\prime}_{\psi}(\phi-\psi)=
\rho^{\prime}(\hat{\bm\omega}\cdot\hat{\bf n})
\sin{(\phi-\psi)},\nonumber\\\nonumber\\&&
\rho^{\prime}_{\psi}(\phi^{\prime}-\psi)=
\rho^{\prime}(\hat{\bm\omega}^{\prime}\cdot\hat{\bf n})
\sin{(\phi^{\prime}-\psi)},\nonumber\\\nonumber\\&&
\partial_{\beta}\left(\hat{\bm\omega}\cdot\hat{\bf n}\right)=
\sum_{\alpha}\omega_{\alpha}\partial_{\beta}n_{\alpha}=
\sin{(\phi-\psi)}\partial_{\beta}\psi,\nonumber\\\nonumber\\&&
\partial_{\gamma}\left(\hat{\bm\omega}^{\prime}\cdot\hat{\bf n}\right)=
\sum_{\delta}\omega_{\delta}^{\prime}\partial_{\gamma}n_{\delta}=
\sin{(\phi^{\prime}-\psi)}\partial_{\gamma}\psi,
\end{eqnarray*}
and therefore
\begin{eqnarray}
\rho^{\prime}_{\psi}(\phi-\psi)\partial_{\beta}\psi&=&
\rho^{\prime}(\hat{\bm\omega}\cdot\hat{\bf n})
\sin{(\phi-\psi)}\partial_{\beta}\psi\nonumber\\\nonumber\\&=&
\rho^{\prime}(\hat{\bm\omega}\cdot\hat{\bf n})
\sum_{\alpha}\omega_{\alpha}\partial_{\beta}n_{\alpha},\nonumber\\\nonumber\\
\rho^{\prime}_{\psi}(\phi^{\prime}-\psi)\partial_{\gamma}\psi&=&
\rho^{\prime}(\hat{\bm\omega}^{\prime}\cdot\hat{\bf n})
\sin{(\phi^{\prime}-\psi)}\partial_{\gamma}\psi\nonumber\\\nonumber\\&=&
\rho^{\prime}(\hat{\bm\omega}^{\prime}\cdot\hat{\bf n})
\sum_{\delta}\omega_{\delta}^{\prime}\partial_{\gamma}n_{\delta},
\end{eqnarray}
so that in terms of the gradients of the nematic director:
\begin{eqnarray}
f_{\rm el}({\bm r})&=&-\frac{\Psi_{\rm exc}(\eta_0)}{8\eta_0}
\sum_{\alpha\beta\gamma\delta}\int\!\!\int d\phi d\phi^{\prime}
{\cal V}_{\beta\gamma}(\phi,\phi^{\prime})\nonumber\\\nonumber\\
&\times&\rho^{\prime}(\hat{\bm\omega}\cdot\hat{\bf n})
\rho^{\prime}(\hat{\bm\omega}^{\prime}\cdot\hat{\bf n})
\omega_{\alpha}\omega_{\delta}^{\prime}
\partial_{\beta}n_{\alpha}
\partial_{\gamma}n_{\delta}.
\end{eqnarray}
Since the direct correlation function of our model is
\begin{eqnarray}
c({\bf r},\hat{\bm\omega},\hat{\bm\omega}^{\prime})=-
\frac{\Psi_{\rm exc}(\eta_0)}{2\eta_0}
v_{\hbox{\tiny exc}}({\bf r},\hat{\bm\omega},\hat{\bm\omega}^{\prime}),
\end{eqnarray}
our expressions for $k_1, k_3$ coincide with the general ones by
Poniewierski and Stecki \cite{Ponie} using a direct correlation function route.

Now in 2D the Frank elastic free energy contains only splay and bend distortions:
\begin{eqnarray}
f_{\rm el}({\bm r})=\frac{1}{2}k_1\left(\nabla\cdot\hat{\bm n}\right)^2+
\frac{1}{2}k_3\left|\nabla\times{\bm n}\right|^2
\end{eqnarray}
Splay and bend are given, respectively, by the deformations:
\begin{eqnarray}
&&\left(\nabla\cdot\hat{\bm n}\right)^2=\left(\frac{\partial n_x}{\partial x}+
\frac{\partial n_y}{\partial y}\right)^2\nonumber\\\nonumber\\&&=\left(\partial_xn_x\right)^2+
\left(\partial_yn_y\right)^2+2\left(\partial_xn_x\right)\left(\partial_yn_y\right),\nonumber\\\nonumber\\
&&\left|\nabla\times\hat{\bm n}\right|^2=\left(\frac{\partial n_y}{\partial x}-
\frac{\partial n_x}{\partial y}\right)^2\nonumber\\\nonumber\\&&=\left(\partial_xn_y\right)^2+
\left(\partial_yn_x\right)^2-2\left(\partial_xn_y\right)\left(\partial_yn_x\right),
\end{eqnarray}
and therefore:
\begin{eqnarray}
&&k_1=-\frac{\Psi_{\rm exc}(\eta_0)}{4\eta_0}
\int\!\!\int d\phi d\phi^{\prime}
\rho^{\prime}(\phi)
{\cal V}_{xx}(\phi,\phi^{\prime})
\rho^{\prime}(\phi^{\prime}),\nonumber\\\nonumber\\
&&k_3=-\frac{\Psi_{\rm exc}(\eta_0)}{4\eta_0}
\int\!\!\int d\phi d\phi^{\prime}
\rho^{\prime}(\phi)
{\cal V}_{yy}(\phi,\phi^{\prime})\rho^{\prime}(\phi^{\prime}).
\nonumber\\
\end{eqnarray}
Here it is assumed that the director goes along the $x$ axis.
\end{document}